\documentclass[%
 reprint,
 amsmath,amssymb,
 aps,
prl
]{revtex4-2}

\usepackage{graphicx}
\usepackage{dcolumn}
\usepackage{bm}

\begin{document}

\preprint{APS/123-QED}

\title{Deterministic Ground-State Search in a Spatial Photonic Ising Machine\\ by Phase Retrieval}

\author{Suguru Shimomura}
 \email{s-shimomura@ist.osaka-u.ac.jp}

\author{Jun Tanida}%
\author{Yusuke Ogura}%
 \affiliation{%
 Graduate School of Information Science and Technology, The University of Osaka, Osaka 565-0871, Japan}%

\date{\today}

\begin{abstract}
A spatial photonic Ising machine (SPIM) solves large-scale combinatorial optimization problems by computing the Ising Hamiltonian through an optical Fourier transform. However, the ground-state search relies on annealing, in which spins are optimized stochastically and sequentially. 
We propose a ground-state search based on phase retrieval (PR), in which all spins are updated simultaneously and deterministically. Imposing an amplitude constraint in the Fourier plane guides the modulated phase distribution, which encodes the spin configuration, toward the optimal solution.
We numerically demonstrate that the proposed scheme reaches the ground state of rank-one Ising Hamiltonians with $10^4$ spins in a single iteration for all trials.
Moreover, a radial rearrangement of the amplitudes and an appropriately designed target pattern alleviate search stagnation and improve the optimization of spin configurations.
The collective and deterministic spin update by phase retrieval enables a fast ground-state search for large-scale combinatorial optimization.
\end{abstract}

\maketitle
Ground-state search in the Ising model provides solutions to combinatorial optimization problems. 
Ising machines are specialized systems designed to search for optimal solutions by mapping these problems to the Ising formulation~\cite{Lucas2014,Mohseni2022}. 
Various Ising machines have been proposed through physical implementations including superconducting circuits~\cite{Johnson2011}, CMOS circuits~\cite{Yamaoka2016}, trapped ions~\cite{Kim2010}, nanomagnets~\cite{Sutton2017}, light propagation~\cite{Ogura2026,Prabhakar2022,Strinati2026}, and optical parametric oscillators~\cite{Inagaki2016,Honjo2021}.
A spatial photonic Ising machine (SPIM) enables the computation of the Ising Hamiltonian through spatial light modulation and an optical Fourier transform~\cite{Pierangeli2019}. 
By exploiting the parallelism of light propagation, Ising Hamiltonians with all-to-all interactions can be computed in constant time independent of the number of spins. 
Several studies have extended the programmability of the SPIM through optical multiplexing techniques including time~\cite{Wang2024}, space~\cite{Sakabe2023,Veraldi2025,Shimomura2025}, and wavelength~\cite{Luo2023,Luo2026}. 
Moreover, phase encoding of the spin configuration and interaction matrix further extends its programmability ~\cite{Fang2021,Sakellariou2025,Sun2022,Ouyang2024}.
Beyond combinatorial optimization, SPIMs have also been extended to statistical learning and applied to image generation~\cite{Yamashita2023,Luo2026}.
Despite this scalability and programmability, reaching the ground or low-energy state of the Ising Hamiltonian remains challenging. 
The optimization performance depends on the search dynamics, which typically employ simulated annealing (SA) with stochastic spin flips over repeated iterations~\cite{Kirkpatrick1983}.
Convergence to the global optimum is established only in the asymptotic limit under a logarithmic cooling schedule, and the number of iterations for spin updates grows rapidly with the problem size~\cite{Bertsimas1993}.
However, the architecture of the SPIM allows the search to employ a deterministic optimization scheme.
In the SPIM, the ground-state search is equivalent to finding the modulated phase distribution that reproduces the target pattern in the Fourier plane. 
Designing the modulated phase pattern is a well-known inverse problem in computer-generated holography~\cite{Lohmann1967} and is structurally analogous to searching for the spin configuration that minimizes the Ising Hamiltonian.
The amplitude constraint in the Fourier plane drives the entire phase distribution toward the configuration that minimizes the Ising Hamiltonian, and a collective update of the spin configuration is realized.

In this Letter, we propose a deterministic optimization framework for the SPIM based on phase retrieval (PR), which reconstructs the modulation pattern from the amplitude constraint imposed in the Fourier plane.
The entire phase distribution is updated at once, and the Ising energy decreases without stochastic sampling.
The search toward the ground state is governed by the convergence of the optical field rather than by the number of spins, enabling a solution to be obtained within a small number of iterations.
Among phase-retrieval strategies~\cite{Fienup1982,Bauschke2002,Elser2018}, we employ the error-reduction scheme based on alternating projection because of its simplicity and monotonic error reduction without residual feedback.
We numerically demonstrate that the proposed scheme reaches the ground state of Ising Hamiltonians with rank-one interaction matrices.

\begin{figure}[tb]
  \centering
  \includegraphics[width = 0.95\linewidth]{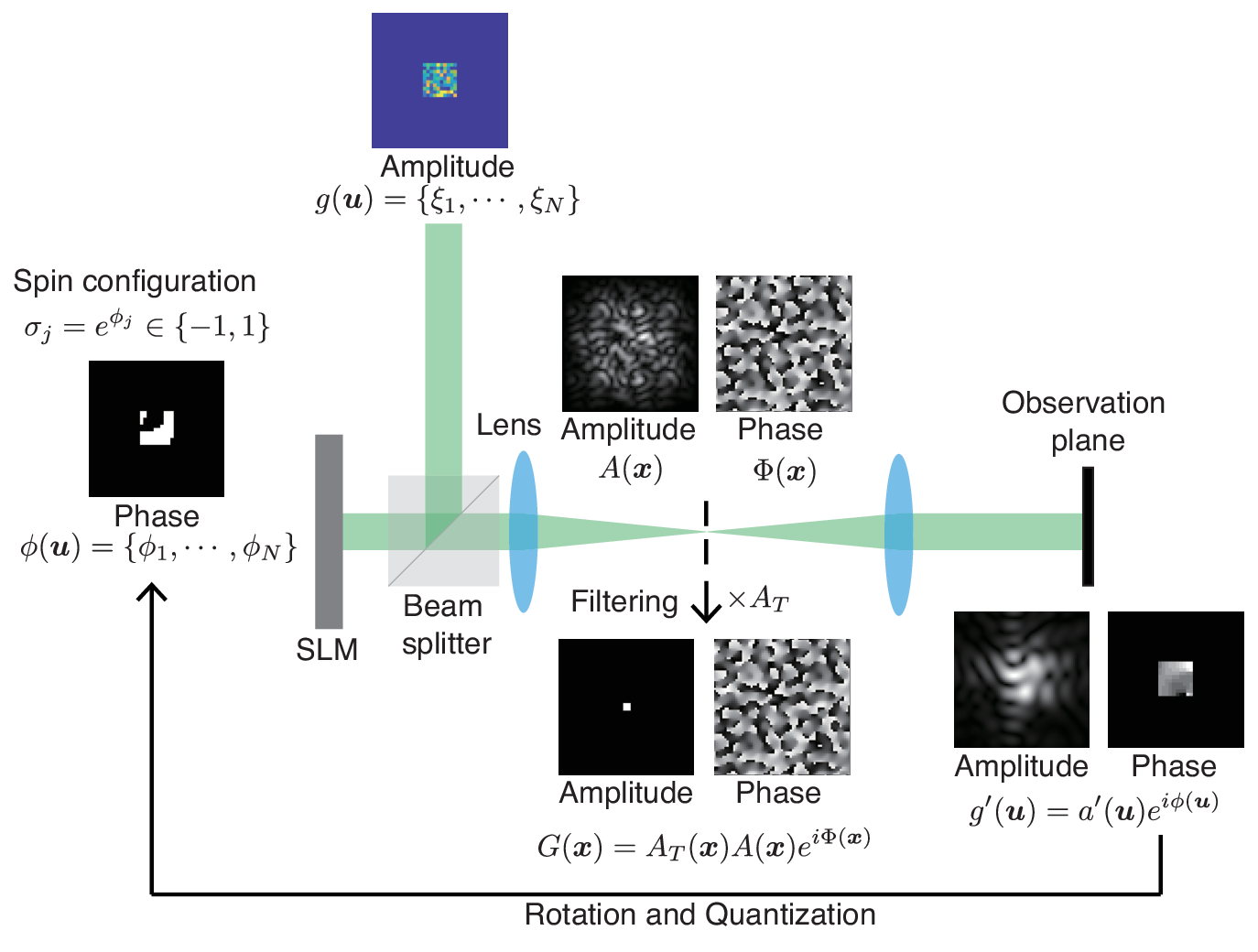}
  \caption{
  Schematic diagram of the PR-based ground-state search for the SPIM. The optical field $G_m$ in the Fourier plane is filtered by the target mask $A_T$, and the filtered field is transformed by the second lens. The phase distribution is extracted, quantized, and displayed on the SLM to generate the next optical field $g_{m+1}$.}
  \label{fig:concept}
\end{figure}

Figure~\ref{fig:concept} shows the schematic diagram of the PR-based ground-state search for the SPIM.
Light with wavelength $\lambda$ and amplitude distribution $\xi=\{\xi_1,\xi_2,\cdots,\xi_N\}$ is incident on a spatial light modulator (SLM), where $N$ is the number of spins, and the SLM plane is divided into $N$ macropixels of width $W$. The phase of light is modulated to encode the binary phase configuration $\phi=\{\phi_1,\phi_2,\cdots,\phi_N\}$, and the spin variables are represented as $\sigma_j=e^{i\phi_j} \in \{-1,1\}$. 
We denote the spatial coordinate on the SLM plane by $\bm{u}=(u,v)$. Then, the center of the $j$th macropixel is $\bm{u}_j=(m_jW,n_jW)$, where $m_j$ and $n_j$ are integers. 
The modulated optical field is $g(\bm{u})=a(\bm{u})e^{i\phi(\bm{u})}=\sum_j\xi_j\sigma_j R(u-Wm_j)R(v-Wn_j)$, where $R$ is the normalized rectangular function of width $W$.
After passing through a lens of focal length $f$, $g(\bm{u})$ is optically Fourier-transformed, and the optical field at the Fourier plane is
\begin{equation}
\begin{aligned}
G(\bm{x})
&=\mathcal{F}[g(\bm{u})]
=A(\bm{x})e^{i\Phi(\bm{x})} \\
&=\frac{1}{\lambda f}\mathcal{S}(\bm{x})\sum_j\xi_j\sigma_j\exp\left(-\frac{2i\pi}{\lambda f}\bm{x}\cdot\bm{u}_j\right),
\end{aligned}
\label{eq:field}
\end{equation}
where $\bm{x}=(x,y)$ is the position in the Fourier plane, and $\mathcal{S}(\bm{x})={\rm sinc}(Wx/\lambda f){\rm sinc}(Wy/\lambda f)$.
The objective in the SPIM is to match $I(\bm{x})=|G(\bm{x})|^2$ to a target intensity $I_T(\bm{x})$.
The evaluation function is
\begin{equation}
\begin{aligned}
\mathcal{O}[\sigma] &= \int I_T(\bm{x}) \cdot I(\bm{x})\, d\bm{x}
\propto \sum_{j,k} J_{jk} \sigma_j \sigma_k, \\
\qquad J_{jk} &= \xi_j \xi_k\, \tilde{C}(\bm{u}_j-\bm{u}_k),
\label{eq:overlap}
\end{aligned}
\end{equation}
where $\tilde{C}=\mathcal{F}^{-1}[I_T\mathcal{S}^2]$. Maximizing $\mathcal{O}$ is equivalent to minimizing the Ising Hamiltonian $H=-\sum_{j,k}J_{jk}\sigma_j\sigma_k$. 
For a point-like intensity target $I_T(\bm{x})=\delta(\bm{x})$ at the optical axis, the coupling matrix is represented as $J_{jk}\propto\xi_j\xi_k$, which is the Mattis model~\cite{Mattis1976}.
Conventional SPIMs search for the optimal spin configuration by measuring the intensity $I(\bm{x})$ iteratively and updating the modulation phase distribution stochastically using SA. 
In our scheme, the phases are allowed to take quantized values, and the complex amplitude of the optical field is used. 
The optical field is filtered by a binary target mask placed at the Fourier plane, $G'(\bm{x})=A_T(\bm{x})G(\bm{x})$, where $A_T(\bm{x})$ represents the target amplitude corresponding to the target intensity $I_T(\bm{x})$.
By passing through another lens, the processed complex amplitude is measured as $g'(-\bm{u})$. Its phase distribution is rotated by 180$^\circ$, quantized, and displayed on the SLM.
Finally, the complex amplitude at the SLM is $g(\bm{u})=a(\bm{u})e^{i\phi'(\bm{u})}$.
At the $m$th iteration, PR yields the input field $g_m(\bm{u})$ at the SLM plane and the field $G_m(\bm{x})$ in the Fourier plane.
Each iteration monotonically reduces the difference between the Fourier-plane field and the target. We define the Fourier-plane error as the optical power outside the target pattern,
\begin{equation}
\begin{aligned}
\mathcal{E}_m^2 &= \int \left|G_m(\bm{x})-A_T(\bm{x})G_m(\bm{x})\right|^2
\,d\bm{x}.
\end{aligned}
\label{eq:error}
\end{equation}
The filtering at the Fourier plane and the amplitude replacement at the SLM plane make the minimum changes required to satisfy the constraints in each domain~~\cite{Fienup1982}. Among the fields with the amplitude distribution $\xi$ and the fields supported on $A_T$, $g_{m+1}$ and $A_TG_{m+1}$ are selected as the nearest to the obtained field $g'_m$ and $G_{m+1}$, respectively. Both $g_m$ and $G'_m$ belong to the respective sets, and Parseval's theorem then gives
\begin{equation}
\begin{aligned}
\mathcal{E}_{m+1}&=\| G_{m+1} - A_T G_{m+1} \|\leq\| G_{m+1} - G'_m \| \\
&=\| g_{m+1} - g'_m \|\leq\| g_m - g'_m \|=\mathcal{E}_m .
\end{aligned}
\label{eq:proj}
\end{equation}
Therefore, the error decreases with each iteration, and the intensity $I_m$ converges toward the target distribution $|A_T|^2=I_T$. 
Note that the total optical power is conserved, and the decrease in $\mathcal{E}_m$ is equivalent to a decrease of the Ising energy $H$.
After the iterations, the phase distribution $\phi^\ast$ that gives the smallest $\mathcal{E}$ is binarized as $\sigma_j = \mathrm{sgn}[\cos \phi_j^\ast]$, which projects the analog phase values onto the spin space $\sigma\in\{-1,+1\}$.
The proposed scheme updates the entire spin configuration simultaneously and deterministically by using the complex amplitude in the Fourier plane.

\begin{figure*}[tb]
  \centering
  \includegraphics[width = 0.95\textwidth]{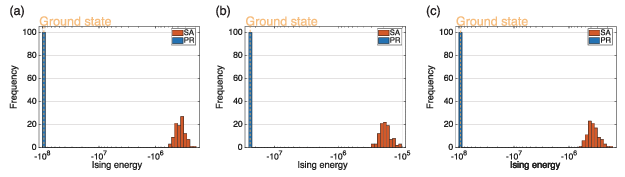}
  \caption{Histograms of the Ising energy with (a) all-to-all coupling $J_{jk}=1$, (b) unweighted random coupling $J_{jk}\in\{0,1\}$, and (c) signed random coupling $J_{jk}\in\{-1,1\}$. Histograms were obtained by the proposed scheme (PR) and SA after 100 iterations.
  In all cases, the Ising energy obtained by PR converged to the ground-state energy. Orange dashed lines indicate the ground-state energy. }
  \label{fig:result1}
\end{figure*}

To demonstrate the convergence toward the ground state, we applied the proposed scheme to problems with rank-one interaction matrices, which are described by the Mattis model and employed as benchmarks for the SPIM ~\cite{Pierangeli2019,Fang2021,Shimomura2025,Ouyang2024}. 
For a rank-one interaction matrix $J_{jk}=\pm\xi_j\xi_k$, the Ising Hamiltonian is described as $H=\mp(\sum_j\xi_j\sigma_j)^2$, and the objective is to maximize or minimize the intensity at the optical axis. 

In the problems maximizing the intensity corresponding to $J_{jk}=\xi_j\xi_k$, the target amplitude pattern was set to a point-like distribution at the optical axis $A_T(\bm{x})=\delta_{\bm{x},0}$, and the number of spins was set to $N = 10^4$. 
To compare the convergence with that of SA, we employed three coupling types defined by $J_{jk}=\xi_j\xi_k$: all-to-all ($\xi_j=1$), unweighted random ($\xi_j\in\{0,1\}$), and signed random ($\xi_j\in\{-1,1\}$).
For the unweighted and signed random cases, each $\xi_j$ was assigned either of its two allowed values with equal probability. For negative $\xi_j$ in the signed random case, $|\xi_j|$ was used as the amplitude, and the corresponding spin was flipped after the iterations and phase binarization.
The initial phase distribution $\phi_0$ was randomly generated from a uniform distribution over $[-\pi,\pi]$. The obtained phase $\phi'$ was quantized to eight levels at each iteration.
The simulation was performed for 100 trials with different random initial phase distributions.
For SA, spins were updated by the Metropolis--Hastings algorithm. The temperature was kept low and constant during the iterations because the energy landscape of these problems has no local minima, and a cooling schedule is not required. The number of spins flipped in each iteration was optimized by a grid search for each problem.
Figure~\ref{fig:result1} shows the histograms of the Ising energy for each graph after 100 iterations. 
For the all-to-all coupling [Fig.~\ref{fig:result1}(a)], the Ising energies obtained by SA remained above the ground-state energy after 100 iterations. In contrast, the proposed scheme reached the ground-state energy in a single iteration for all 100 trials.
The same behavior was observed for the unweighted random coupling [Fig.~\ref{fig:result1}(b)] and the signed random coupling [Fig.~\ref{fig:result1}(c)]. 
For these couplings, the energy landscape of the rank-one interaction has no local minima, and the alternating projection can reach the global solution in a single iteration. 
These results demonstrate that the proposed scheme reaches the ground state of the Ising energy deterministically.

\begin{figure*}[t]
  \centering
  \includegraphics[width = 0.95\textwidth]{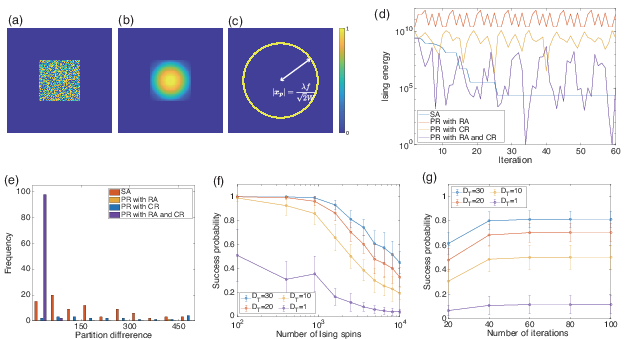}
  \caption{Amplitude distributions $\xi$ for (a) the conventional arrangement and (b) the radial arrangement. (c) Designed target pattern $A_T$ consisting of a circular ring passing through the Nyquist points. The ring width was set to the main-lobe width of the diffraction pattern of the full modulation area. (d) Evolution of the Ising energy for each optimization scheme (RA, radial rearrangement; CR, circular ring target). (e) Histograms of partition differences for 2.5$\times10^3$ spins (100 trials). (f) Success probability as a function of the number of spins after 60 iterations. (g) Success probability as a function of the number of iterations for 2,500 spins. 
  The values in (f) and (g) are averaged over 10 problem instances with 100 trials each. The error bars denote the standard deviation across the instances.}
  \label{fig:result2}
\end{figure*}

Next, we evaluated the performance of the proposed scheme by employing number-partitioning problems, which have local minima in the Ising energy landscape and whose interaction matrix is described by $J_{jk} = -\xi_j\xi_k$. 
In these problems, a set of numbers $\xi$ is partitioned into two subsets such that the sums of the two subsets are as nearly equal as possible, and the target amplitude is set to $A_T(\bm{x})=1-\delta_{\bm{x},0}$.
Under this constraint, the Fourier-plane error is the on-axis intensity alone. Since $I(\bm{0})$ is only a small part of the total intensity, the reconstructed field $g'$ barely differs from $g$; the iteration reaches a fixed point, and the Ising energy stagnates~\cite{Fienup1986}.
To alleviate the stagnation, we designed the spatial arrangement of the amplitude $\xi$ and the target pattern $A_T$. 
First, we changed the assignment of the amplitudes to the macropixels. 
In the conventional SPIM, $\xi_j$ is assigned to the $j$th macropixel in index order [Fig.~\ref{fig:result2}(a)]. In the proposed scheme, we sorted and assigned $\xi_j$ in the order of the radial distance of the macropixels from the optical axis [Fig.~\ref{fig:result2}(b)].
When the number of spins is large and the distribution $\xi$ can be regarded as a continuous function, the radial rearrangement of the amplitudes results in nearly equal amplitudes for the $j$th and $\bar{j}$th macropixels, $\xi_{\bar{j}}\approx\xi_j$, where $\bm{u}_{\bar{j}}=-\bm{u}_j$. 
Therefore, the amplitude distribution becomes nearly symmetric with respect to the optical axis.
Note that the assignment of $\xi_j$ to the macropixel positions can be chosen freely without changing the problem since the Ising energy depends only on the set $\xi$.
Defining the symmetric and antisymmetric spin components as $\sigma_j^{\pm}=(\sigma_j\pm\sigma_{\bar{j}})/2$, Eq.~(\ref{eq:field}) is approximated as
\begin{equation}
\begin{aligned}
G(\bm{x})\simeq
\frac{2}{\lambda f}\mathcal{S}(\bm{x})
\sum_{\langle j,\bar{j}\rangle}\xi_j
\bigg[&
\sigma_j^{+}\cos\left(\frac{2\pi}{\lambda f}\bm{x}\cdot\bm{u}_j\right) \\
&-i\sigma_j^{-}\sin\left(\frac{2\pi}{\lambda f}\bm{x}\cdot\bm{u}_j\right)
\bigg],
\end{aligned}
\label{eq:sym}
\end{equation}
where the summation runs over the point-symmetric macropixel pairs $\langle j,\bar{j}\rangle$.
Second, we restricted the target amplitude to the four diagonal Nyquist points of the macropixel array, $A_T(\bm{x})=\sum_{p=1}^{4}\delta_{\bm{x}-\bm{x}_p,0}$ with $\bm{x}_p=(\pm\lambda f/(2W),\pm\lambda f/(2W))$, which are located at the radial distance $|\bm{x}_p|=\lambda f/(\sqrt{2}W)$ from the optical axis.
At the diagonal Nyquist points, the phase in Eq.~(\ref{eq:sym}) becomes $2\pi\bm{x}_p\cdot\bm{u}_j/(\lambda f)=(\pm m_j\pm n_j)\pi$, and the intensities become $I(\bm{x}_p) \propto \bigg|\sum_{j}\kappa_j\xi_j\sigma_j\bigg|^2$, where $\kappa_j=(-1)^{m_j+n_j}$. Minimizing $\mathcal{E}^2$ is equivalent to maximizing the intensity $I(\bm{x}_p)$, and the minimum condition of $\mathcal{E}^2$ is obtained by aligning all the phasors $\kappa_j\xi_j\sigma_j$ to a common sign as
$\sigma_j=\pm\,\kappa_j\,\mathrm{sgn}(\xi_j)$, and the maximum intensity is $I_{\max}(\bm{x}_p)\propto{\rm sinc}^4(1/2)\big(\sum_j|\xi_j|\big)^2$.
At $\bm{x}=\bm{0}$, the intensity is  
\begin{equation}
\begin{aligned}
I(\bm{0})
&=\bigg(\sum_{j}\kappa_j|\xi_j|\bigg)^2 \\
&=\bigg(\sum_{j\in \alpha}|\xi_j|-\sum_{j\in \beta}|\xi_j|\bigg)^2,
\end{aligned}
\label{eq:partition}
\end{equation}
where $\alpha$ and $\beta$ denote the two subsets with $\kappa_j=+1$ and $\kappa_j=-1$, respectively.
As a result, the spin configuration maximizing the Nyquist-point intensity partitions the set $\{|\xi_j|\}$ into the two subsets.
Because the radial rearrangement assigns nearly equal amplitudes to adjacent macropixels belonging to the opposite positions, the two subset sums nearly balance and $I(\bm{0})\approx0$.
Therefore, maximizing $I(\bm{x}_p)$ transfers the optical power from the optical axis to the Nyquist points and drives the Ising energy toward its minimum. 
In practice, the radial rearrangement equalizes the paired amplitudes only approximately, $\xi_{\bar{j}}\approx\xi_j$. The minimum condition is not met exactly, and optical power remains outside the four points [Fig.~\ref{fig:result2} (c)]. 
To reduce the effect of the residual asymmetry while still concentrating the optical power around the Nyquist radius, we designed the target amplitude as a circular ring of radius $x_{\rm N}$ passing through the four points. 
The width of the ring was set to $w=2\lambda f/W_{\rm total}$, which is equal to the main-lobe width of the diffraction pattern of the full modulation area, where $W_{\rm total}$ is the side length of the square amplitude distribution $\xi$. 
Because only a small region of the Fourier plane is used for the reconstruction, the Fourier-plane error $\mathcal{E}$ becomes large, and the field $g'$ differs substantially from $g$. Therefore, a single iteration can update many spins, and the search continues to progress instead of stagnating.
For each problem instance, the amplitudes $\xi_j$ were sampled uniformly from $\{1,2,\cdots,\lfloor N-\frac{1}{2}\log_2(\pi N/6)\rfloor\}$.
Figure~\ref{fig:result2} (d) shows the evolution of the Ising energy $H=I(\bm{0})$ during the iterations for $N=2.5\times10^3$. 
While the energy decreased monotonically with SA, PR with only the radial rearrangement (RA) or only the circular ring (CR) target stagnated. 
In contrast, the proposed scheme with both RA and the CR target reached lower energies than those obtained by SA. 
This result demonstrates that the proposed scheme searches for the ground state while escaping from local minima. 
Figure~\ref{fig:result2} (e) shows histograms of the partition difference $D=|\sum_{j\in\alpha}\xi_j-\sum_{j\in\beta}\xi_j|$ after 60 iterations for $N=2.5\times10^3$.
Without RA, the PR search stagnated at larger partition differences than those obtained by SA. By employing RA and the CR target, the proposed scheme reached lower-energy states and found perfect partitions $D\leq1$ in 17 of 100 trials.
To evaluate the optimization accuracy, we defined the success probability $P_{\rm suc}$ as the proportion of trials with $D\leq D_T\in\{1,10,20,30\}$. 
For each condition, 10 problem instances (amplitude sets $\xi$) were generated independently, and
100 trials with different random initial phase distributions were performed for each number of spins.
Figure~\ref{fig:result2} (f) shows $P_{\mathrm{suc}}$ as a function of the number of spins for the proposed method with RA and the CR target after 60 iterations. Even as the number of spins $N$ increased, approximate solutions were found in some trials, and perfect partitions were obtained for each $N$. 
Overall, amplitude rearrangement and target-mask design alleviate search stagnation, and both approximate and optimal solutions could be found in far fewer iterations than the number of spins. 
We also estimated the number of iterations required for convergence. Figure~\ref{fig:result2} (g) shows the success probability as a function of the number of iterations for $2.5\times10^3$ spins. 
The probability saturated after 60 iterations because, even with the proposed scheme, convergence to the global optimum is not guaranteed, and the search can stagnate at a local minimum. 
However, the ground state could be reached within a few iterations in some trials; therefore, reinitializing the phase distribution could be effective when the search stagnates.

In annealing-based Ising machines, the ground-state search requires a number of iterations that grows with the number of spins. 
Since the proposed scheme updates the entire spin configuration at once, the number of iterations is independent of the number of spins. This advantage becomes more pronounced as the number of spins increases.
In the SPIM, the time to solution and the energy consumption grow with the number of iterations even when a high-speed SLM and a low-power laser are employed~~\cite{Veraldi2025}.
The reduced iteration count leads to a fast and energy-efficient ground-state search.
In addition, the proposed scheme could be combined with SPIM techniques such as spatial multiplexing and phase encoding, which extend the class of Ising Hamiltonians that can be implemented.
To obtain the complex amplitude $g'(\bm{u})$ at the SLM plane, it is necessary to measure the phase distribution $\phi'(\bm{u})$. The phase information can be reconstructed from the intensity
distribution through interference with a reference beam~~\cite{Latychevskaia2019}. 
By combining the proposed scheme with off-axis holography, the ground-state search can be performed by capturing interference patterns with a camera and displaying the reconstructed phase distribution on the SLM.

In conclusion, we proposed a collective ground-state search method for the SPIM based on PR, in which the amplitude constraint on the optical field deterministically updates the entire spin configuration at once.
We numerically demonstrated that the scheme reaches the ground state of rank-one Ising Hamiltonians with $10^4$ spins in a single iteration. Moreover, amplitude rearrangement and target-pattern design extend the search to nonconvex energy landscapes, such as those encountered in number partitioning.
These results suggest that search dynamics based on PR offer collective updates of spins and rapid optimization in optical Ising machines.

This research was supported by the JST-ALCA-Next Program (Grant Number JPMJAN23F2) and JSPS KAKENHI (Grant Number 25H01884).

\bibliography{Mycollection}

\end{document}